# Unit cell orientation of tetragonal-like BiFeO$_3$ thin films grown on highly miscut LaAlO$_3$ substrates

C. Beekman,[1] W. Siemons,[1] T.Z. Ward,[1] J.D. Budai,[1] J.Z. Tischler,[2] R. Xu,[2] W. Liu,[2] N. Balke,[3] J. H. Nam,[1,4] and H.M. Christen[1]

[1]Materials Science and Technology Division, Oak Ridge National Laboratory, Oak Ridge, TN, 37831, USA
[2]Advanced Photon Source, Argonne National Laboratory, Argonne, IL, 60439, USA.
[3]Center for Nanophase Materials Sciences, Oak Ridge National Laboratory, Oak Ridge, TN, 37831, USA
[4]Optic and Electronic Ceramics Division, Korea Institute of Ceramic Engineering and Technology (KICET), Seoul 153-801, Republic of Korea

**Abstract**
Synchrotron and lab-scale x-ray diffraction shows that tetragonal-like T'-BiFeO$_3$ films on miscut LaAlO$_3$ substrates ($\alpha < 5°$) exhibit (00l)-planes tilted away from those of the substrate as predicted by the "Nagai model" (except for miscut <0.2°). Tilts as large as 1° are achieved even in 100 nm thick films, strikingly larger than those observed in other perovskites. We attribute this to the large *c/a* ratio and the high crystalline coherency of the T'-BiFeO$_3$/LaAlO$_3$ interface. This coherency is possible through an observed "diagonal-on-diagonal" alignment. Interestingly, the substrate miscut does not influence the relative population of monoclinic domains.

**Main text**

Strain engineering, i.e. the intentional modification of crystal structure, microstructure, and morphology of epitaxial thin films via the mismatch between the film's and the substrate's lattice parameter and symmetry, is a powerful tool to tune functional properties of materials.[1-3] Similarly, the presence of step edges on the substrate surface resulting from substrate miscut can also have a large impact on the microstructure of thin films, for example by breaking the four-fold surface symmetry of a (001)-oriented cubic substrate. Thus, for a film consisting of unit cells having a non-square in-plane lattice, the degeneracy between different in-plane film-substrate alignments may be lifted in an approach referred to as domain engineering.[4,5] Furthermore, a surface step locally results in a three-dimensional clamping of the film material instead of the biaxial stress experienced elsewhere. As a consequence of the resulting non-uniform strain in the out-of-plane direction, the (00l) planes of the film may become tilted with respect to those of the substrate. For the case of $Ga_xIn_{1-x}As$ on GaAs, Nagai demonstrated[6] that simple geometrical arguments explain the observed tilting of the film planes by an angle $\gamma$ when grown on a substrate having a miscut angle $\alpha$, as shown in Fig. 1. In this simple model, the film out-of-plane lattice parameter is constrained to match the substrate at each step edge and then gradually relaxes to its unstrained value across the flat surface terrace. In fact, with out-of-plane lattice parameters for the film and substrate, $c_F$ and $c_S$ respectively, one expects $\tan\gamma / \tan\alpha = (c_F - c_S)/c_S$, and thus a negative (positive) tilt if $c_F > c_S$ ($c_F < c_S$).

Observing a tilt corresponding to this "Nagai model" is possible only in a system where the clamping in three dimensions is strong, i.e. where the film/substrate interface shows a low defect density, and only if competing mechanisms, such as misfit dislocations with inclined Burgers vectors,[7] do not dominate. Quantitative verification of the Nagai model has therefore remained elusive in oxide heteroepitaxy except for very thin films having a lattice parameter similar to that of the substrate (for example,[8] 15 nm of $SrRuO_3$ on 4°-miscut $SrTiO_3$, resulting in a tilt of ~0.06°). A much larger film/substrate tilt should be observed in a system with $c_F >> c_S$ (or $c_F << c_S$). Interestingly, if for example $c_F > c_S$, $c_F$ must also be significantly larger than the substrate's in-plane ($a_S$) lattice parameter in order to avoid accommodation of the vertical stress by formation of a-c domain twinning, i.e. the rotation of the long axis of the film unit cell into the plane of the substrate, as observed for example in the case of $PbTiO_3$ on $SrTiO_3$.[9]

In this study, we test the Nagai model using the highly-stained ("tetragonal-like")[10,11] T' polymorph of $BiFeO_3$ grown on $LaAlO_3$ substrates that are miscut with respect to the [001] direction with $0 < \alpha < 5°$ (pseudocubic notation is used throughout for the description of $LaAlO_3$, with $a_S$ = 3.789 Å). Due to its multiferroic properties, $BiFeO_3$ has been studied extensively in film form and is known to undergo a strain-induced phase transition from the mildly-strained R' polymorph[12,13] on $SrTiO_3$ substrates to the T' form[10,11,14-16] on $LaAlO_3$ or $YAlO_3$ substrates. The T' polymorph on $LaAlO_3$ is monoclinic with $a_F$ = 3.81(2) Å, $b_F$ = 3.73(2) Å, $c_F$ = 4.67(2) Å and $\beta$ = 88.1 (3)°.[15] Its in-plane lattice area $a_F \cdot b_F$ = 14.2(2) Å$^2$ is thus a very close match to the in-plane lattice of $LaAlO_3$ ($a_S^2$ = 14.36 Å$^2$). We show that the film's in-plane alignment is of diagonal-on-diagonal type (i.e., $BiFeO_3$ [110] || $LaAlO_3$ [110]), which results in an almost perfect film-substrate lattice match ($a_F \cdot b_F / a_S^2$ = 0.99(1)). As a consequence, even films with thicknesses of 100 nm exhibit a low defect density; film-substrate tilts as large as 1° are observed that quantitatively follow the Nagai model within measurement accuracy.

The experiments were performed on films grown by pulsed laser deposition (PLD) on substrates with miscuts of <0.2°, 0.6°, 4.5° and 5°. The films were grown in a 50 mTorr oxygen background pressure while the substrates were kept at a temperature of 675 °C. A pulsed KrF excimer laser with a wavelength of 248 nm was focused on a 10% excess Bi $BiFeO_3$ sintered pellet with an energy density of 0.4 J/cm$^2$ and operated at 2 Hz, resulting in an average deposition rate of ~0.03 Å/pulse. Synchrotron x-ray microdiffraction was performed at beamline 34-ID-E at the Advanced Photon Source at Argonne National Laboratory[17] (0.5 μm beam diameter at a 45° incidence angle). Atomic force microscope (AFM) images were taken on a Veeco D3100 operated in tapping mode, the piezoresponse force microscopy (PFM) measurements were performed on a Veeco Dimension Nanoscope 5 AFM equipped with external lock-in amplifiers, and x-ray diffraction (XRD) reciprocal space mapping (RSM) was performed on a laboratory PanAlytical X'Pert thin film diffractometer with Cu Kα radiation.

The T' BiFeO$_3$ films grown on LaAlO$_3$ substrates with $\alpha$ < 0.2° show the typical mixed phase morphology with a striped phase embedded within the monoclinic T' phase (see Fig. 2a).[11,18-20] The stripes correspond to an additional polymorph of BiFeO$_3$ that reversibly form upon cooling (i.e., they are not structural defects).[21] Because they constitute a small volume fraction of the sample and exhibit a significantly different lattice parameter than the T' polymorph, they can safely be ignored for the present study. However, within the flat T' regions, different monoclinic domains coexist (as is evidenced in the PFM data of Fig. 2b and additional XRD data shown as supplemental material[22]), a point to which we return below. As expected for the growth of strained films, use of a substrate with a large miscut results in a very different morphology due to stepbunching.[23].

In Fig. 3 we show typical x-ray intensity $I(2\theta, \Delta\omega)$ maps (where $\Delta\omega = \omega - \theta$) containing the 002$_{pc}$ film and substrate peaks of the same film for two orthogonal in-plane ($\varphi$) orientations (i.e. parallel and perpendicular to the step edges of the substrate). Perpendicular to the step edges we observe a large tilt of the film planes by 1.0(1)° (Fig. 3 a); parallel to the step edges tilting is absent (Fig. 3 b). Note that the *c*-axis lattice parameter of the film remains unchanged when substrate miscut is increased. The tilt angles extracted from $I(2\theta, \Delta\omega)$ maps for films grown on substrates with $\alpha$ = 0.6°, 4.5°, and 5° and with different film thicknesses (data points for thicknesses ranging from 40 to 100 nm overlap) are presented in Fig. 4 (an AFM image of a 40 nm BiFeO$_3$ film grown on 4.5° miscut LaAlO$_3$ is shown in the inset, with step heights of ~ 20 unit cells). The figure shows a remarkable agreement between the measured tilt angles and the predictions from the Nagai model (indicated by the solid line), which for the above-indicated lattice parameters simplifies to $\gamma \approx 0.225\alpha$ at small $\alpha$. Note that this is different from previous work on R' BiFeO$_3$ on SrRuO$_3$-buffered miscut SrTiO$_3$[5,24]: there, the SrRuO$_3$-tilt with respect to the SrTiO$_3$ substrate follows the Nagai model as mentioned above, but BiFeO$_3$ tilts in a direction opposite to that predicted by the Nagai model, as has been attributed to defect formation. This agrees with the observation that for BiFeO$_3$ on strained SrRuO$_3$, the in-plane lattice match $a_F^2 / a_S^2$ = 1.03(1) is much poorer than our present value of $a_F b_F / a_S^2$ = 0.99(1) for BiFeO$_3$/LaAlO$_3$ as discussed above.

The data points at $\alpha$ < 0.2° shown in Fig. 4 were obtained by synchrotron microdiffraction on a 50 nm thick film as described below. At these low miscuts, the terrace widths are large (> 250 unit cells, and larger than the film thickness), and while tilting and strong localized distortions presumably still occur near the step edges, the majority of the film has its (00l) planes parallel to those of the substrate.

Next we investigate the orientation of the in-plane lattice vectors of the BiFeO$_3$ film with respect to those of the substrate. Synchrotron x-ray microdiffraction was performed on a 50 nm film grown on a substrate with $\alpha$ < 0.2°. We first note that the rhombohedral structure of the LaAlO$_3$ results in the formation of twins, and hence the different substrate domains necessarily exhibit different amounts of miscut. Therefore, by using microdiffraction, regions with different values of $\alpha$ can be distinguished (0 < $\alpha$ < 0.2°), leading to the data points for $\alpha$ < 0.2° mentioned above (Fig. 4). Laue diffraction patterns were recorded with a wide x-ray energy range (7 – 30 keV) and while scanning the 0.5 µm beam across multiple substrate domains (see schematic drawing and optical micrograph in Figs. 5a and b, respectively). The red and green dashed lines in Fig. 5b indicate the domain boundaries and are a guide to the eye. Fig. 5c shows four representative snapshots of the same section of the Laue diffraction pattern around the {2 0 13} LaAlO$_3$ peaks (indicated by the label "L") taken at different locations on the film (for a full movie see supplemental material).[22] The splitting of the LaAlO$_3$ reflection into two peaks confirms the multidomain character of the substrate; it does not vary from location to location over short distances due to the penetration of the x-ray beam across multiple domains (see schematic in Fig. 5a). Each pattern shows two {108} BiFeO$_3$ peaks labeled "B", for a total of four BiFeO$_3$ domain variant types in this general region of the sample, which indicates a twinning of the film within each LaAlO$_3$ domain. Since a pair of BiFeO$_3$ {108} peaks is observed in all Laue patterns, nanoscale twinning exists with a length scale smaller than the 0.5 µm beam size. Measurement of the spacing between BiFeO$_3$ peaks determines that the *a*- and *b*-axes of the film are rotated away from the in-plane substrate *a*- and *b*-axes by 0.8(2)°. Given the typical value for *a/b* = 1.021(8),[15] a diagonal-on-diagonal alignment as sketched in Fig. 6 results in a predicted rotation of 0.6(2)°. The results thus show that the film aligns in-plane along the face diagonals of the LaAlO$_3$ (BiFeO$_3$ [110] || LaAlO$_3$ [110]$_{pc}$), as is often observed in epitaxial perovskites.[25] Macroscopic x-ray diffraction through the (022), (2-22), (202) and (222) reflections of a 100 nm thick BiFeO$_3$ film on a 4.5° miscut LaAlO$_3$ substrate, as the sample is rotated about the Φ-axis, show a doublet along the projection of the substrate's <h00> direction and a triplet along the <hh0> direction respectively (data not shown).[26] This confirms that the same type of twinning (i.e. diagonal-on-diagonal alignment) occurs on all samples regardless of substrate miscut. Furthermore, the same splitting is observed for the (022) reflection of a 18 nm thick BiFeO$_3$ film grown on <0.2° LaAlO$_3$ (i.e. a film that is too thin to contain the striped phase), which clearly shows that the striped phase is not responsible for the observed splitting in the {108} BiFeO$_3$ peaks. Note that the two film spots in each of the frames of Fig. 5c correspond to two domains as drawn in Fig. 6; an additional splitting resulting from a monoclinic "shear" distortion along the ±*a* direction remains unresolved in these Laue patterns but results in the piezoelectric domain pattern visible in the PFM image of Fig. 2b. During a scan across the sample, a change between patterns (1) and (2) [or (3) and (4)] corresponds to a change between the multidomain structure of Fig. 6 to its mirror image, and this occurs repeatedly before the beam moves to the next LaAlO$_3$ domain. Therefore, the length scale over which the pattern of Fig. 6 changes to its mirror image is smaller than the LaAlO$_3$ domain size (~10 μm), yet is larger than the x-ray microbeam (~0.5μm), and is significantly larger than the ferroelastic domains corresponding to the monoclinic shear distortion (Fig. 2b). Inspection of RSMs taken for samples with different miscut α indicate that neither the *a/b* ratio, the monoclinic angle, nor the relative abundance of the ferroelectric domains is affected by the substrate miscut. For our T'-BiFeO$_3$ films, therefore, substrate miscut does not lead to the same type of domain engineering as reported for R'-BiFeO$_3$, where a preferential ferroelectric polarization direction is induced as a consequence of substrate steps.[5]

To summarize, our study on T'-BiFeO$_3$ films on miscut LaAlO$_3$ substrates (0 < α < 5°) shows that the BiFeO$_3$ (00l)-planes tilt away from those of the substrate as predicted by geometrical arguments in the Nagai model (except at the smallest miscuts, α < 0.2°, where the substrate terrace step width greatly exceeds our film thickness). Tilts as large as 1° are achieved even in 100 nm thick films; these values are strikingly larger than those earlier observed in other perovskite materials. We attribute this to the unique *c/a* ratio of T'-BiFeO$_3$, which is large enough to result in a strong effect and to make the formation of *a*-domains energetically unfavorable, and to a high crystalline coherency of the T'-BiFeO$_3$/LaAlO$_3$ interface. This coherency is possible through a "diagonal-on-diagonal" alignment between the film and the substrate that leads to a very small effective lattice mismatch between the T'-polymorph of BiFeO$_3$ and LaAlO$_3$. Contrary to R'-BiFeO$_3$ on SrTiO$_3$, the relative population of different monoclinic structural domains remains unaffected by the miscut (even at α = 4.5°).

**Acknowledgements**
Research supported by the U. S. Department of Energy (DOE), Basic Energy Sciences (BES), Materials Science and Engineering Division (HMC, JDB, WS, CB, TWZ, and JHN). Piezoresponse force microscopy (NB) was conducted at the Center for Nanophase Materials Sciences, Oak Ridge National Laboratory, and x-ray microdiffraction (JZT, RX, WL) at the Advanced Photon Source, Argonne National Laboratory, both supported by the Scientific User Facilities Division of DOE-BES.

Captions

Figure 1. Diagram explaining the Nagai[6] model showing the negative tilt angle $\gamma$ of the (00l) film planes with respect to those of the substrate when the film is grown on a substrate with miscut $\alpha$ (with $c_S$ and $c_F$ the lattice parameters of the substrate and film respectively and L the length of the terraces of the miscut substrate). The film surface is parallel to the substrate surface.

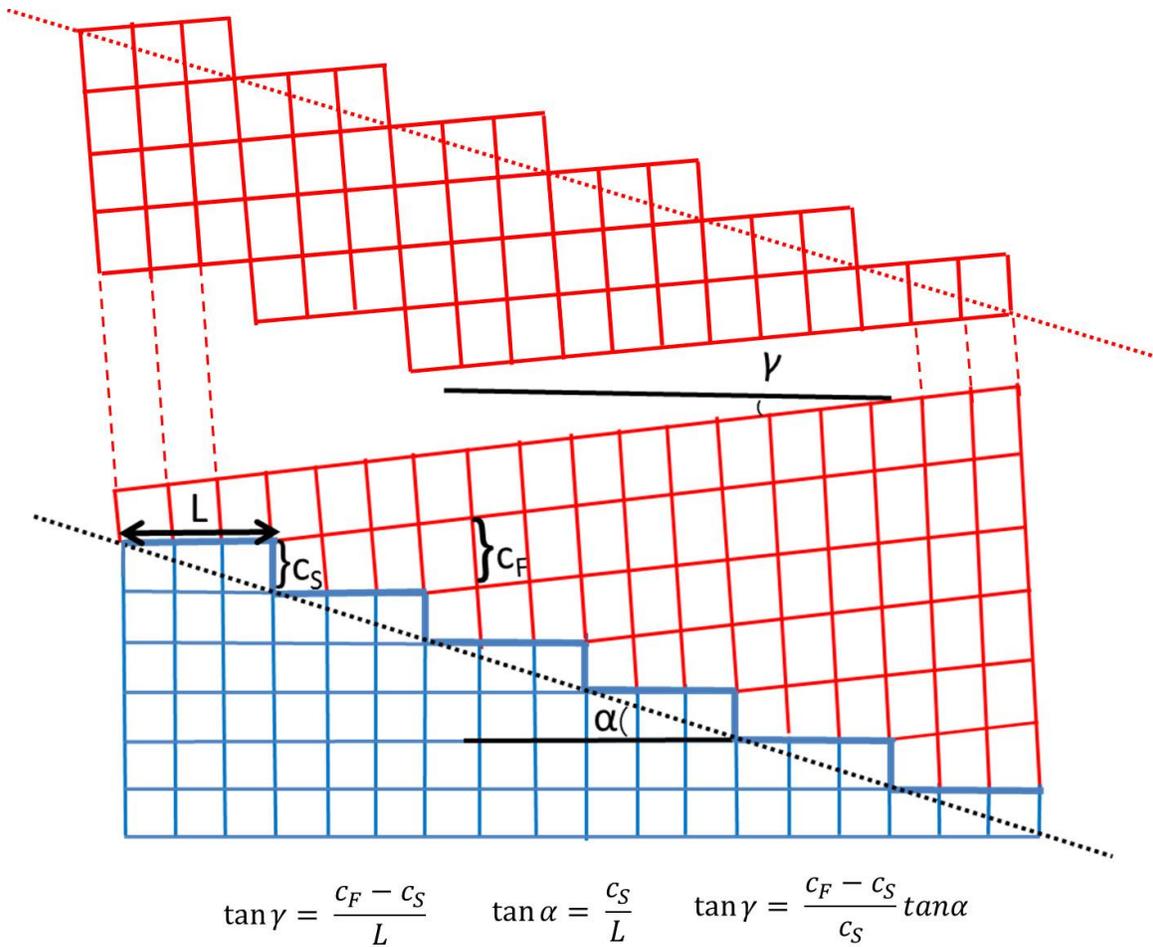

$$\tan\gamma = \frac{c_F - c_S}{L} \qquad \tan\alpha = \frac{c_S}{L} \qquad \tan\gamma = \frac{c_F - c_S}{c_S}\tan\alpha$$

Figure 2. a) AFM image at room temperature. Scan size is 1 μm x 1 μm. b) In-plane PFM amplitude image corresponding to the AFM topography shown in a.

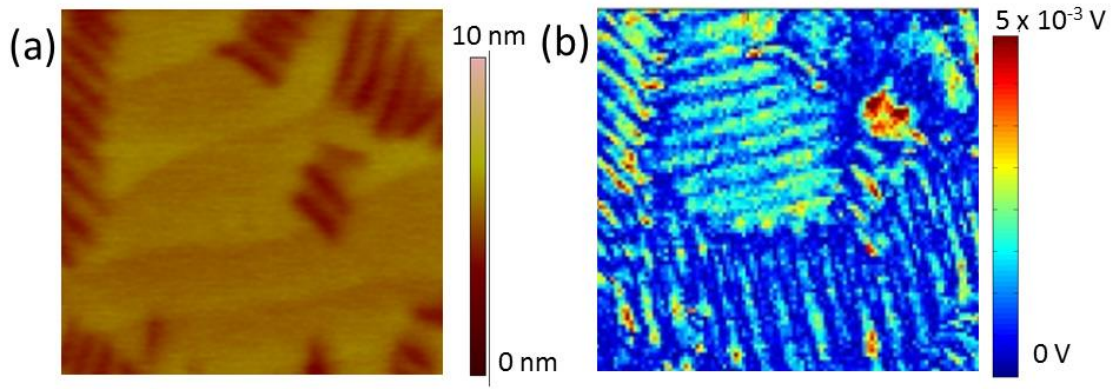

Figure 3. X-ray intensity $I(2\theta, \Delta\omega)$ maps of the $002_{pc}$ diffraction peak for a 40 nm thick $BiFeO_3$ film grown on a $LaAlO_3$ substrate with 4.5° miscut: (a) perpendicular to the step edges of the substrate (along the miscut vector) and (b) along the step edges of the substrate. The substrate peak is at $2\theta = 47.96°$ and the film peak is at $2\theta = 38.7°$, the dashed red lines show that for different in-plane orientations of the sample the peaks occur at the same values for $2\theta$. The vertical axis shows the amount with which the film (00l) planes are tilted away from the substrate (00l) planes.

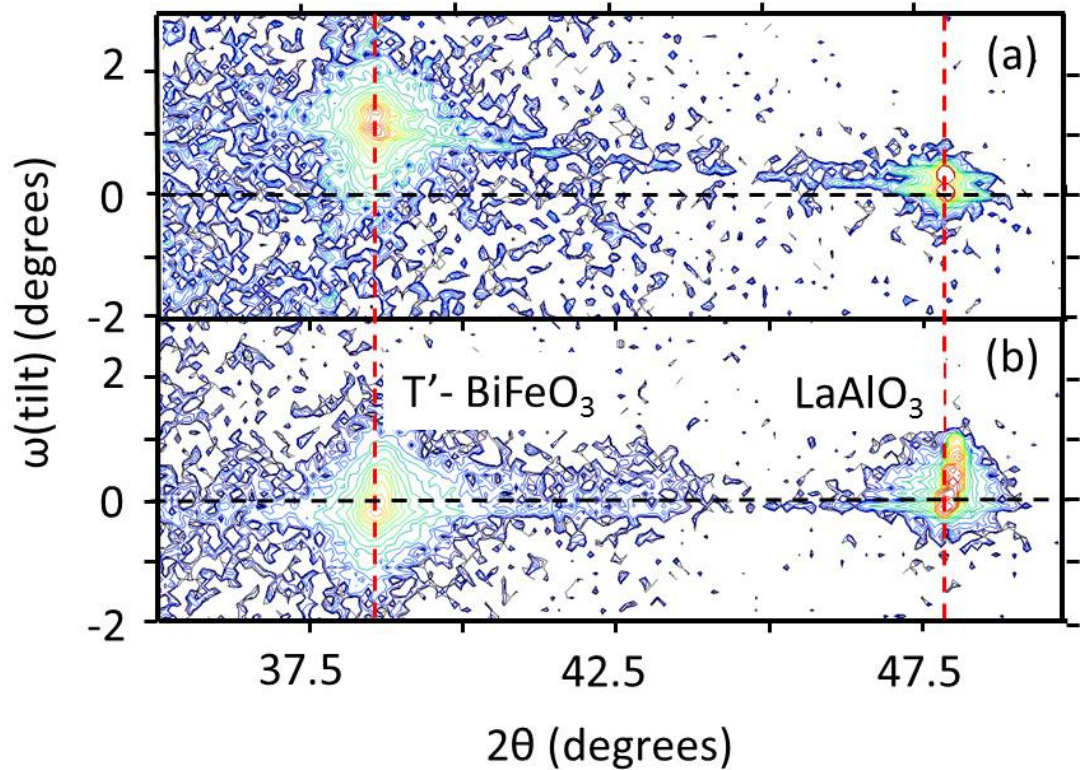

Figure 4. The amount of tilt of the (00l) film planes as extracted from the synchrotron microdiffraction experiments (films for miscut < 0.2°) and from X-ray diffraction (films on highly miscut LaAlO$_3$). Symbols overlap for films of different thicknesses between 40 and 100 nm. The red line represents the tilt angles predicted by the geometrical arguments proposed by Nagai.[6] Inset: AFM image of BiFeO$_3$ films grown on 4.5° LaAlO$_3$ substrates (scan size: 4 x 4 µm); stepbunching is observed with an average step height of about 10 nm.

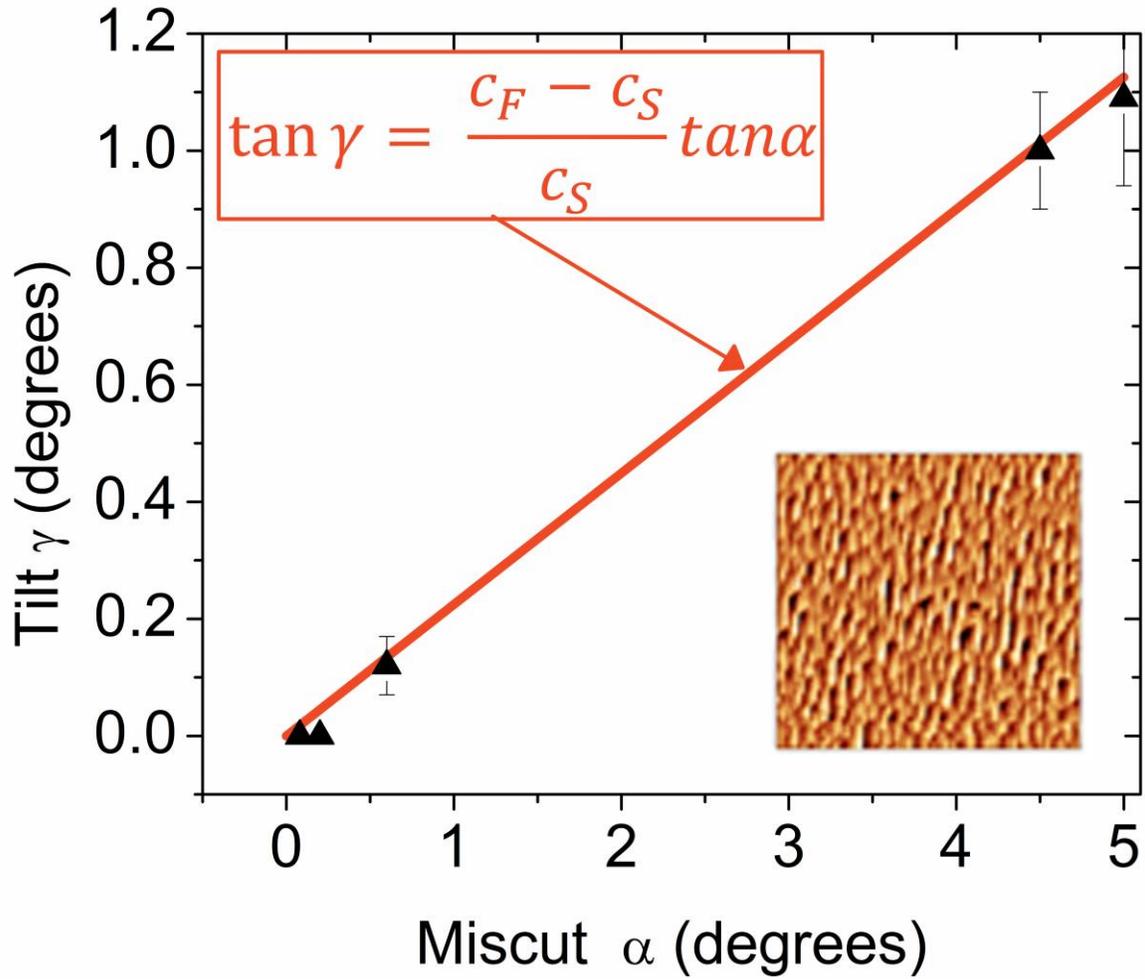

Figure 5. Synchrotron microdiffraction, while scanning across the multiple domains of the substrate. (a) diagram of the substrate twinning (blue) with the BiFeO$_3$ film (red). b) Optical microscope image showing the scanning direction and distance with the LaAlO$_3$ domain boundaries indicated by the dashed lines. A movie of the Laue diffraction pattern, while scanning across the different LaAlO$_3$ domains (indicated by the arrow) is shown in the supplemental material (Fig. S2).[22] (c) Section of the Laue diffraction patterns showing the {2 0 13}$_{pc}$ LaAlO$_3$ peaks (indicated by L) and the {1 0 8}$_t$ BiFeO$_3$ peaks (indicated by B) at four different locations. The pairs (1,2) and (3,4) are taken within one LaAlO$_3$ domain, respectively.

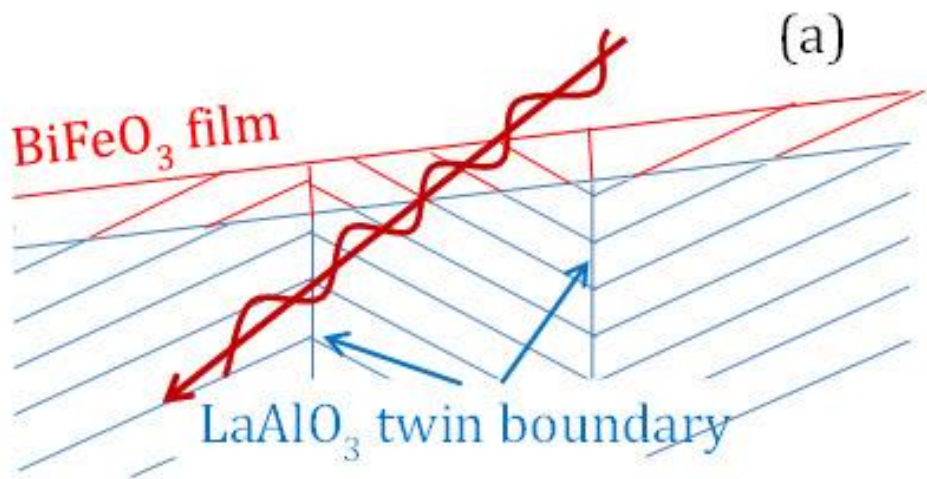
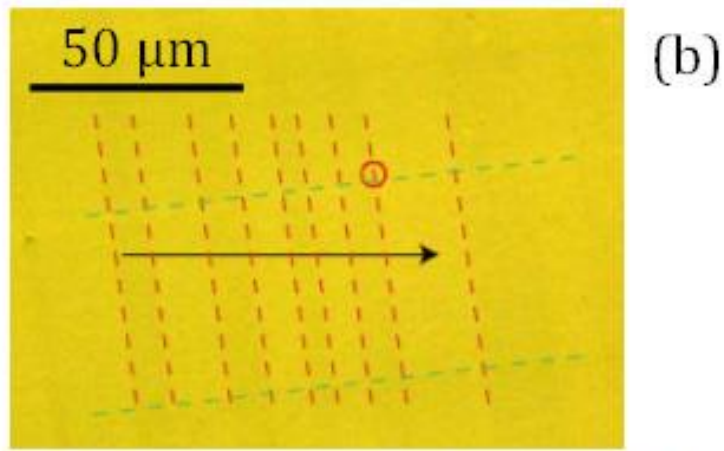
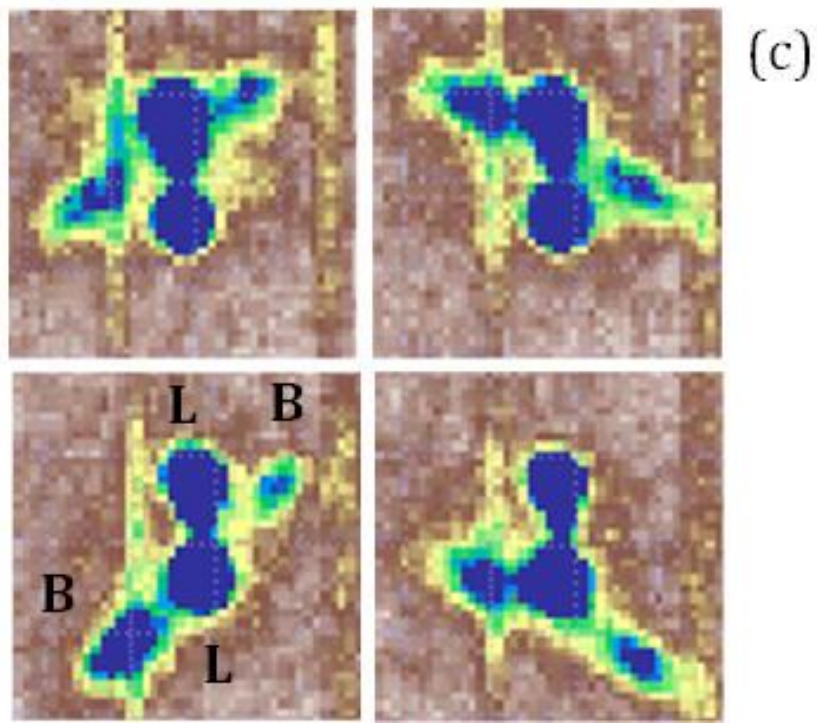

Figure 6: Schematic drawing of the in-plane BiFeO$_3$ lattice (red) tilted away from the LaAlO$_3$ lattice (blue) showing the creation of BiFeO$_3$ twins. The dashed line indicates the twin boundary.

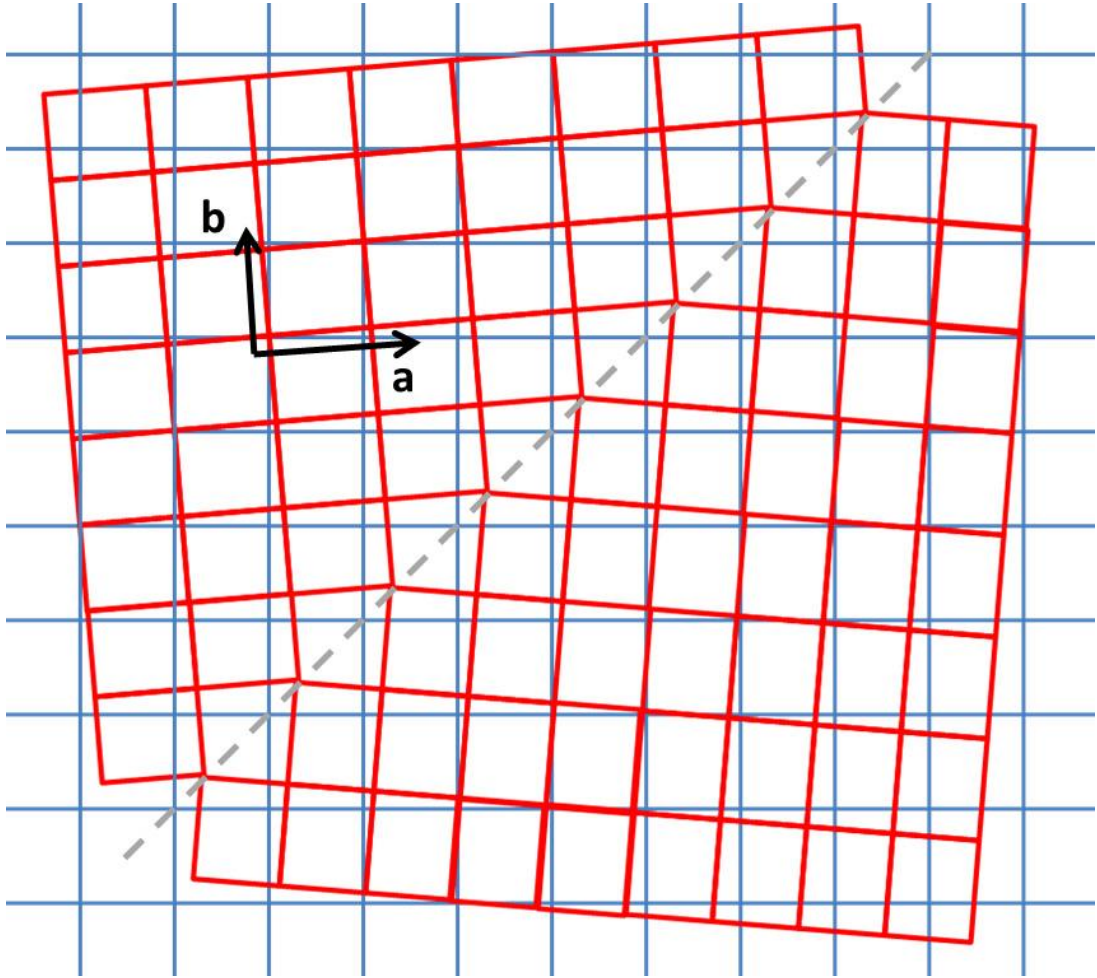